\documentclass[aip,jcp,superscriptaddress,reprint,twocolumn]{revtex4-1}
\usepackage{epsfig}
\usepackage{amsmath}
\usepackage{amsfonts}
\usepackage{graphicx}
\usepackage{fancybox}
\usepackage{subfigure}
\usepackage{color}
\usepackage[titletoc]{appendix}
\usepackage{sidecap}
\usepackage[mathscr]{euscript}

\newcommand{\erf}{\ensuremath{\mathop{\rm erf}}}
\newcommand{\erfc}{\ensuremath{\mathop{\rm erfc}}}
\newcommand{\rb}{\mathbf{r}}
\newcommand{\pb}{\mathbf{p}}

\newcommand{\Rb}{\mathbf{R}}

\newcommand{\avg}[1]{\left<#1\right>}
\newcommand{\len}[1]{\left|#1\right|}
\newcommand{\brac}[1]{\left[#1\right]}

\newcommand{\para}[1]{\left(#1\right)}

\newcommand{\etal}{\emph{et al.}}

\newcommand{\rhoqs}{\ensuremath{\rho^{q\sigma}}}
\newcommand{\rhoq}{\ensuremath{\rho^q}}

\newcommand{\rhoG}{\ensuremath{\rho_G}}

\newcommand{\phiRl}{\ensuremath{\phi_{\rm R1}}}

\newcommand{\V}{\ensuremath{\mathcal{V}}}
\newcommand{\Vr}{\ensuremath{\mathcal{V}_{\rm R}}}

\newcommand{\Vrl}{\ensuremath{\mathcal{V}_{\rm R1}}}

\newcommand{\PhiRl}{\ensuremath{\Phi_{\rm R1}}}

\newcommand{\kT}{\ensuremath{k_{\rm B}T}}

\newcommand{\Hb}{\ensuremath{\mathcal{H}}}
\newcommand{\Zb}{\ensuremath{\mathcal{Z}}}
\newcommand{\Rbar}{{\textbf{R}}}
\newcommand{\Pbar}{{\textbf{P}}}

\begin{document}




\title{Modeling Nuclear Quantum Effects on Long Range Electrostatics in Nonuniform Fluids}


\author{Richard C. Remsing}
\affiliation{Department of Chemistry and Chemical Biology,
Rutgers University, Piscataway, NJ 08854}

\begin{abstract}
Nuclear quantum effects play critical roles in a variety of molecular processes,
especially in systems that contain hydrogen and other light nuclei, such as water. 
For water at ambient conditions, nuclear quantum effects are often interpreted as
local effects resulting from a smearing of the hydrogen atom distribution.
However, the orientational structure of water at interfaces determines long range effects
like electrostatics through the O-H bond ordering that is impacted by nuclear quantum effects. 
In this work, I examine nuclear quantum effects on long range electrostatics of water
confined between hydrophobic walls using path integral simulations. 
To do so, I combine concepts from local molecular field (LMF) theory with path integral methods
at varying levels of approximation to develop an efficient and physically intuitive approaches
for describing long range electrostatics in nonuniform quantum systems. 
Using these approaches, I show that quantum water requires larger electrostatic forces
to achieve the same level of interfacial screening as the corresponding classical system. 
This work highlights subtleties of electrostatics in nonuniform classical and quantum molecular systems,
and the methods presented here are expected to be of use to efficiently model
nuclear quantum effects in large systems.
\end{abstract}




\maketitle


\raggedbottom

\section{Introduction}

The quantum mechanical behavior of atomic nuclei can have important
effects on molecular processes. 
Chemical kinetics, isotope effects, and the structure of molecular systems
is influenced by nuclear quantum effects (NQEs). 
Therefore, NQEs must be adequately described in theoretical or computational
models of molecular systems, especially when light nuclei or low temperatures are
involved, because NQEs are prevalent at these conditions. 
In water at ambient conditions, NQEs primarily impact hydrogen atoms
and affect the structure, dynamics, and chemical reactivity of liquid water~\cite{markland2018nuclear,ceriotti2016nuclear,marsalek2017quantum,wang2014quantum,marx2000solvated,tuckerman2002nature,xu2021importance,Raugei:JACS:2003,tang2021nuclear,morrone2008nuclear,ceriotti2013nuclear,marx2000solvated,zhuang2019many,paesani2010nuclear,spura2015nuclear,ojha2018nuclear,litman2018decisive}. 
Quantum mechanical treatment of water is often considered
to smear out the positions of the hydrogens, although non-trivial electronic quantum effects can couple to the
nuclear fluctuations in ab initio treatments. 
In all of these situations, NQEs are considered mainly local effects,
impacting the hydrogen bond structure and dynamics of water.
However, any changes in the hydrogen bond structure can also impact non-local properties of water,
especially in nonuniform systems where changes in liquid structure are magnified by interfaces.
In particular, water preferentially adopts specific orientations at interfaces~\cite{lee1984structure,lee1994comparison,LCW,Chandler:2005aa,Remsing:JPCB:2013,jamadagni2011hydrophobicity,giovambattista2012computational,rego2022understanding}. 
These orientational preferences control interfacial electrostatics and are dictated by the water hydrogen bond network.
This hydrogen bond network, in turn, is sensitive to NQEs.
Ultimately, this suggests that NQEs can indirectly impact orientational properties of water in nonuniform
systems and the consequent electrostatic properties of the system. 
Here, I investigate NQEs on the long range electrostatic properties of water in nonuniform systems
using molecular simulations and quantum statistical mechanics combined with liquid state theory.
A standard approach to modeling NQEs
is path integral molecular dynamics (PIMD) simulations~\cite{Parrinello_1984,Tuckerman_1996,ceperley1995path}, as well as
the closely related ring polymer molecular dynamics (RPMD) simulation approach~\cite{Habershon:2013aa}. 
In these approaches, each quantum particle is replaced by an isomorphic classical
\emph{ring polymer} composed of $P$ beads (or monomers). 
For large enough $P$, one can determine exact statistical averages of static quantities. 
Using RPMD, one can model approximate quantum dynamics by evolving the system
in time using classical dynamics, but in the extended phase space of the ring polymer system~\cite{Habershon:2013aa}.
Modeling the ring polymer system amounts to simulating $P$ coupled copies or replicas
of the classical system, enabling the use and extension of standard computational algorithms.
Replicating the system $P$ times leads to significant computational cost, especially
when $P$ is large. 
This is especially true for long range interactions, like electrostatics, which
are already expensive in the classical system ($P=1$). 
However, by judiciously separating electrostatic interactions into short and long range components,
the number of beads needed to evaluate the long range part can be significantly reduced, even
down to the $P=1$ limit~\cite{MARKLAND2008256,Markland:2008}. 
This approach --- ring polymer contraction (RPC) --- evaluates electrostatic interactions on
a subset of beads and distributes the result over all the beads of the polymer, reducing
the number of evaluations of long range interactions from $P$ to $P' < P$.
Here, I focus on the limit where long range interactions are evaluated only
at the centroid of each ring polymer $P'=1$. 
While RPC can significantly reduce the computational cost of PIMD and RPMD simulations,
this approach is still plagued with the usual problems associated with evaluating
long range electrostatics --- Ewald summations are expensive and conceptually difficult. 
The conceptual issues are particularly problematic, because Ewald sums and other lattice summation techniques
can often lead to significant geometry-dependent finite size effects, as well as other
artifacts associated with periodic boundary conditions (PBCs)~\cite{figueirido1995finite,hub2014quantifying,Pan2017}. 
An appealing alternative to lattice sums is local molecular field (LMF) theory
and related developments~\cite{LMFDeriv,Remsing:2016ib,Hu2014,SSM,gao2023local}. 
LMF theory relies on a physically intuitive separation of electrostatic interactions into
short and long range components, much like RPC, and replaces all electrostatic interactions
by the short range part. 
The averaged effects of the long range interactions are then modeled
through an effective external field that is chosen to produce the same structure as the
full system with long range interactions. 
Accurate structure, thermodynamics, and even dynamics can be efficiently and accurately predicted by
LMF simulations, making it a useful alternative to lattice summations. 

In this work, I combine LMF theory and the path integral isomorphism to develop RPC-based approximations and model nuclear quantum effects in water confined between model nonpolar walls. 
After a brief review of RPC and LMF theory, I discuss strategies for combining
the two to obtain approaches for the efficient and accurate predictions of water structure
between model hydrophobic walls.
These approaches can aid in reducing the cost of PIMD and RPMD simulations
while also helping to build physical intuition regarding the effects of long range interactions
in heterogeneous quantum systems. 
%

\section{Theory}

\subsection{Ring Polymer Contraction}
In standard PIMD and RPMD simulation approaches, Feynman's path integral
formulation of quantum mechanics is used to model
a system of $N$ distinguishable particles with Hamiltonian $\Hb(\Pbar,\Rbar)$,
where $\Pbar$ and $\Rbar$ represent the momenta and positions of all $N$ particles
at a single point in phase space~\cite{FeynmanPathIntegrals,Chandler:JCP:1981,Habershon:2013aa}. 
The partition function, $\Zb$, of this system can be approximated by that
of an isomorphic system composed of classical ring polymers, each composed of $P$ beads,
\begin{align}
\Zb &\approx \Zb_P = \frac{1}{(2\pi\hbar)^{3NP}} \int d\Pbar \int d\Rbar e^{-\beta\Hb_P(\Pbar,\Rbar)/P},
\end{align}
which becomes an equality in the $P\rightarrow\infty$ limit. 
Here, $\beta=(\kT)^{-1}$ and the ring polymer Hamiltonian is
\begin{align}
\Hb_P(\Pbar,\Rbar)&= \sum_{i=1}^N \sum_{\alpha=1}^P \para{ \frac{\len{\pb_i^{(\alpha)}}^2}{2m_i} + \frac{1}{2} m_i \omega_P^2 \len{\rb_i^{(\alpha)} - \rb_i^{(\alpha-1)}}^2} \nonumber \\
&+\sum_{\alpha=1}^P V(\Rbar^{(\alpha)}),
\end{align}
where the last term is the sum of the potential energy over all beads,
$\rb_i^{(\alpha)}$ and $\pb_i^{(\alpha)}$ are the position and momentum of bead $\alpha$ of site $i$,
$\Rb^{(\alpha)}$ represents the position vector for bead $\alpha$ of all $N$ particles in a single configuration,
$m_i$ is the mass of site $i$, and $\omega_P=P/\beta\hbar$ is the spring constant. 
In general, this Hamiltonian can involve many-body interactions, 
but here I focus on one- and two-body interactions,
\begin{equation}
V(\Rbar^{(\alpha)}) = \sum_{i=1}^N \phi(\rb_i^{(\alpha)}) + \sum_{i=1}^{N-1} \sum_{j=i+1}^N w\para{\len{\rb_i^{(\alpha)} - \rb_j^{(\alpha)}}},
\end{equation}
where the one-body potential $\phi(\rb)$ arises from an external field.
The two-body potential $w(r)$ involves Lennard-Jones (LJ) and electrostatic interactions,
\begin{equation}
w(r_{ij}) = u_{\rm LJ}(r_{ij}) + q_i q_j v(r_{ij}),
\end{equation}
where $r_{ij}=\len{\rb_i-\rb_j}$, $u_{\rm LJ}(r)$ is the LJ potential,
and $v(r)=1/r$. 
Lennard-Jones interactions are typically truncated at some distance and their
effects beyond the cutoff accounted for with a correction. 
In contrast, electrostatic interactions are long-ranged and are typically
evaluated using lattice summation techniques like Ewald summation.
Lattice summations are generally expensive and significantly
increase the cost of PI simulations when the system is replicated $P$ times
to construct the ring polymers. 
Ring polymer contraction (RPC) can be used
to reduce this cost by lowering the number of beads
on which long range interactions need to be evaluated~~\cite{MARKLAND2008256,Markland:2008}. 
RPC splits the interparticle interactions into two components, $V(r)=V_{\rm S}(r) + V_{\rm L}(r)$,
where $V_{\rm S}(r)$ is the short range part of the potential
and $V_{\rm L}(r)$ is the long range part of the potential.
The splitting is chosen such that $V_{\rm L}(r)$ varies slowly on the length scale
of the largest ring polymer, estimated by the ensemble averaged radius of gyration. 
For water at ambient conditions, this length scale is close to the free particle limit,
$\lambda_{\rm free}=\sqrt{\beta\hbar^2/4m}$. 
When the potential varies slowly on the scale of $\lambda_{\rm free}$,
the total interaction between two ring polymers can be approximated
by $P$ times the interaction between their centroids,
\begin{equation}
\sum_{\alpha=1}^P V_{\rm L}\para{\len{\rb_i^{(\alpha)} - \rb_j^{(\alpha)}}} \approx P V_{\rm L}(\len{\bar{\rb}_i - \bar{\rb}_j})
\label{eq:rpc}
\end{equation}
where $\bar{\rb}_i$ is the centroid of ring polymer $i$.
Forces can be readily evaluated following previous work~\cite{MARKLAND2008256,Markland:2008}.
The centroid RPC approximation, Eq.~\ref{eq:rpc}, significantly reduces the cost of evaluating long range
interactions without sacrificing accuracy~\cite{MARKLAND2008256,Markland:2008}. 
Using local molecular field theory, summarized in the next section, we can further
reduce the cost of evaluating long range interactions.

\subsection{Local Molecular Field Theory}
LMF theory accounts for the averaged effects of long range electrostatics
with a renormalized or effective electrostatic potential~\cite{LMFDeriv,Remsing:2016ib}. 
The first step in determining this potential is to separate intermolecular Coulomb interactions
into short and long range components. 
For LMF theory to be valid, long range interactions must vary slowly over typical
nearest neighbor distances. 
As such, LMF theory separates the $1/r$ portion of the Coulomb potential according to
\begin{align}
v(r)&=\frac{1}{r} \\
&= \frac{\erfc(r/\sigma)}{r} + \frac{\erf(r/\sigma)}{r} \\
&\equiv v_0(r) + v_1(r),
\end{align}
where $\sigma$ is the LMF smoothing length that is on the order of intermolecular correlations,
$v_0(r)$ is the short-range component of the electrostatic interactions, and $v_1(r)$ is the long-range component.
For water at ambient conditions, previous work has shown that $\sigma\ge3$~\AA~\cite{MolPhysLMF},
and here I use a conservative value of $\sigma=4.5$~\AA. 
In LMF theory, the full model is replaced by its Gaussian-truncated (GT) counterpart,
in which $v(r)$ is replaced by $v_0(r)$ for all sites in the system. 
The averaged effects of fluid-fluid long range electrostatics are taken into account
through the renormalized electrostatic potential
\begin{align}
\Vr(\rb) &= \V(\rb) + \int d\rb' \rhoq(\rb') v_1(\len{\rb-\rb'}) \\
&\equiv \V(\rb) + \V_{\rm S}(\rb),
\label{eq:lmf}
\end{align}
where $\V(\rb)$ is the external electrostatic potential present in the full system, equal to zero for the systems studied here,
$\rhoq(\rb)$ is the ensemble averaged singlet charge density of the system,
given by
\begin{equation}
\rhoq(\rb)=\avg{\rhoq(\rb;\Rbar)},
\end{equation}
and
$\rhoq(\rb;\Rbar)$ is the charge density operator evaluated in configuration $\Rbar$.
The charge density operator is defined in the next section. 
In general, the external field contains can be split into short and long range parts, $\V(\rb)=\V_0(\rb)+\V_1(\rb)$,
such that the LMF can be rewritten in a way that isolates the long range interactions as $\Vr(\rb) = \V_0(\rb) + \Vrl(\rb)$.
Equation~\ref{eq:lmf} is self-consistent and can be solved
through brute-force simulations or
using methods based in linear response theory~\cite{Hu:PRL:2010}.
Simulating the GT model in the presence of $\Vr(\rb)$ yields structures in agreement
with the full system~\cite{Rodgers:PNAS:2008,baker2020local,cox2020dielectric}, and thermodynamics can be obtained
with previously-derived corrections~\cite{rodgers2009accurate,Remsing:JSP:2011,Remsing:2016ib,Remsing:JTCC:2018,SSM}. 
%

\subsection{Solving the Local Molecular Field Equation for Quantum Systems}

The LMF potential can be obtained by writing the ring polymer
expression for the charge density operator of the system in configuration $\Rbar$ as~\cite{Craig_2004,Habershon:2013aa}
\begin{align}
\rhoq(\rb;\Rbar) &= \frac{1}{P}\sum_{\alpha=1}^P \sum_{i=1}^N q_i^{(\alpha)} \delta\para{\rb-\rb_i^{(\alpha)}(\Rbar)} \\
&=\frac{1}{P} \sum_{\alpha=1}^P \rho^{q\alpha}(\rb;\Rbar).
\end{align}
Using this expression for the charge density, the LMF potential is given by
\begin{align}
\Vr(\rb)&=\V(\rb) + \frac{1}{P} \sum_{\alpha=1}^P \int d\rb' \rho^{q\alpha}(\rb') v_1(\len{\rb-\rb'}) \\
&\equiv \V(\rb) + \frac{1}{P} \sum_{\alpha=1}^P \V_{\rm S}^{(\alpha)}(\rb) 
\label{eq:rplmf}
\end{align}
I will refer to this as the path integral local molecular field (PI-LMF). 
The PI-LMF Equation~\ref{eq:rplmf} must be solved self-consistently. 
A self-consistent solution for $\Vr(\rb)$ can be found by iterating with molecular simulations, 
but this is expensive, especially for quantum systems. 
This can be circumvented by iterating to self-consistency using linear response theory (LRT) instead of simulations
to predict the density induced by each field, as described by Hu and Weeks for classical fluids~\cite{Hu:PRL:2010}.
I now describe how to extend this LRT approach to solving the LMF equation to path integral models.

In a system of quantum particles described within the path integral formalism, the Hamiltonian is replaced by the corresponding
approximation involving ring polymers composed of $P$ beads each. 
Ignoring the momenta --- we are only concerned with configurational averages here --- the path integral Hamiltonian can
be written as
\begin{align}
\Hb_P(\Rbar) &= \frac{1}{P}\sum_{\alpha=1}^P \brac{U_0^{(\alpha)}(\Rbar) + \Phi_0^{(\alpha)}(\Rbar) + \PhiRl^{(\alpha)}(\Rbar) } \\
& = \Hb_{P,0}(\Rbar) + \frac{1}{P} \sum_{\alpha=1}^P \PhiRl^{(\alpha)}(\Rbar),
\end{align}
where the bond potentials between neighboring beads are included in $U_0^{(\alpha)}(\Rbar)$.
The Hamiltonian $\Hb_{P,0}(\Rbar) = \frac{1}{P}\sum_{\alpha=1}^P \brac{U_0^{(\alpha)}(\Rbar) + \Phi_0^{(\alpha)}(\Rbar)}$
represents the purely short ranged (reference) system,
$U_0^{(\alpha)}(\Rbar)$ is the total potential energy of the short range pair interactions for bead $\alpha$, and
$\Phi_0^{(\alpha)}(\Rbar)$ is the corresponding total short range one-body potential energy. 
The total potential energy of the long range interactions for each bead is contained within
$\PhiRl^{(\alpha)}(\Rbar) = \int d\rb \rho^{q\alpha}(\rb;\Rbar) \Vrl(\rb)$.
Using this separation of the Hamiltonian into short and long range components, 
the average charge density can be written as an ensemble average in the short range system according to
\begin{align}
\rhoq(\rb) & = \avg{\frac{1}{P} \sum_{\alpha=1}^P \rho^{q(\alpha)}(\rb;\Rbar)} \nonumber \\
& = \frac{\avg{ \frac{1}{P} \sum_{\alpha=1}^P\rho^{q(\alpha)}(\rb;\Rbar) e^{-\frac{\beta}{P}\sum_{\gamma=1}^P \PhiRl^{(\gamma)}(\Rbar)}}_0} {\avg{e^{-\frac{\beta}{P}\sum_{\alpha=1}^P \PhiRl^{(\alpha)}(\Rbar)}}_0}.
\end{align}
Now, noting that the instantaneous bead-averaged field energy is
\begin{equation}
{\Phi}_{\rm R1}(\Rbar) = \frac{1}{P}\sum_{\alpha=1}^P \PhiRl^{(\alpha)}(\Rbar),
\end{equation}
we can rewrite the charge density as an average over configurations in the short range system,
\begin{equation}
\rhoq(\rb) = \frac{\avg{ \rhoq(\rb;\Rbar) e^{-\beta {\Phi}_{\rm R1}(\Rbar)}}_0}{\avg{e^{-\beta {\Phi}_{\rm R1}(\Rbar)}}_0}.
\end{equation}
We can then linearize this expression for the charge density of quantum particles to obtain the linear response approximation
\begin{equation}
\rhoq(\rb) \approx \avg{\rhoq(\rb;\Rbar)}_0 -\beta\avg{\delta \rhoq(\rb;\Rbar) \delta {\Phi}_{\rm R1}(\Rbar)}_0,
\label{eq:qlrtq}
\end{equation}
where $\delta \rhoq(\rb;\Rbar) = \rhoq(\rb;\Rbar) - \avg{\rhoq(\rb;\Rbar)}_0$
and $\delta {\Phi}_{\rm R1}(\Rbar)=\Phi_{\rm R1}(\Rbar) - \avg{\Phi_{\rm R1}(\Rbar)}_0$.
Equation~\ref{eq:qlrtq} is analogous to the classical result of Hu and Weeks~\cite{Hu:PRL:2010},
except the classical operators are replaced by their bead-averaged counterparts~\cite{Craig_2004,Habershon:2013aa}.
Equations~\ref{eq:rplmf} and~\ref{eq:qlrtq} are the main results of this section and are used to obtain a self-consistent solution to the LMF equation through
iteration. 
The benefits of using Eq.~\ref{eq:qlrtq} instead of the more traditional form of the linear response approximation are magnified
in path integral treatments of quantum systems. 
In this case, the more traditional expression
\begin{equation}
\rhoq(\rb) \approx\avg{\rhoq(\rb;\Rbar)}_0 - \frac{\beta}{P^2} \sum_{\alpha=1}^P \sum_{\gamma=1}^P\int d\rb' \phiRl(\rb')  \chi^{qq}_{\alpha\gamma}(\rb,\rb')
\label{eq:qtradq}
\end{equation}
involves pair correlations between all beads in the system,
including those on different slices of imaginary time ($\alpha$ and $\gamma$),
through the quantum  charge-charge linear response function~\cite{Chandler:JCP:1981,Shinoda_2001,Sese_2002}
\begin{equation}
\chi^{qq}_{\alpha\gamma}(\rb,\rb') = \avg{\delta \rho^{q(\alpha)}(\rb;\Rbar) \delta \rho^{q(\gamma)}(\rb';\Rbar)}.
\end{equation}
In addition to the difficulties of evaluating a six-dimensional correlation function in a nonuniform system,
the need to evaluate correlations between different points in imaginary time further increases the expense of using
Eq.~\ref{eq:qtradq}. 
Because of these difficulties, the much more efficient Eq.~\ref{eq:qlrtq} is preferred to solve
the LMF equation for path integral models.

\subsection{Combining Local Molecular Field Theory and Ring Polymer Contraction}

While the solution to the LMF equation in the previous section
can be obtained from simulation results and linear response theory,
the slowly-varying nature of the long range potentials suggests that simpler approximations can be exploited
to more efficiently solve the LMF equation. 
One approach is to combine LMF theory with RPC.
A fundamental concept in both LMF theory and RPC is the separation of interaction potentials
into short and long range components based on physical principles. 
In RPC, electrostatic interactions are separated so that the long range component
is slowly-varying over the size of the ring polymer~\cite{MARKLAND2008256,Markland:2008}.
In LMF theory, electrostatic interactions are separated so that the long range component
is uniformly slowly-varying over typical nearest-neighbor distances (or a correlation length) in the liquid~\cite{LMFDeriv,Remsing:2016ib}. 
In liquids like water at ambient conditions, the typical separation length scales are similar, and I will indicate this
by the LMF smoothing length $\sigma$. 
I follow these principles and use the typical LMF smoothing length of $\sigma=4.5$~\AA \ to separate
the potential into short and long range components, $v_0(r)$ and $v_1(r)$, respectively,
as described in the previous section. 
In RPC, the electrostatic (pair) potential $v_1(r)$ is evaluated between centroid positions. 
By combining RPC and LMF, long range pair interactions are completely removed, $V_{\rm L}=0$.
The averaged effects of long range interactions are instead accounted for via the effective one-body electrostatic (LMF) potential
$\Vr(\rb)$, and RPC can be used to evaluate the long range part of the LMF potential at centroid positions only;
\begin{equation}
\sum_{i=1}^{N-1} \sum_{j>i}^N P V_{\rm L}(\len{\bar{\rb}_i - \bar{\rb}_j}) \rightarrow \sum_{i=1}^N P\Vrl(\bar{\rb}_i).
\end{equation}
This strategy results in the LMF-RPC scheme for evaluating long range interactions.
To determine the effective field $\Vr(\rb)$ within the LMF-RPC scheme
using Eq.~\ref{eq:qlrtq}, the long range potential energy is evaluated only at the location of the centroid. 
Defining the centroid charge density operator, 
\begin{align}
\bar{\rho}^q(\rb;\Rbar) = \sum_{i=1}^N q_i \delta\para{\rb-\bar{\rb}_i(\Rbar)},
\end{align}
where $\bar{\rb}_i$ is the position of the centroid of particle $i$,
the total long range potential energy within the LMF-RPC approximation is
\begin{equation}
\Phi_{\rm R1}^{\rm RPC}(\Rbar) = \int d\rb \bar{\rho}^q(\rb;\Rbar) \Vrl(\rb).
\label{eq:rpcphi}
\end{equation}
This corresponds to evaluating the field $\Vrl(\rb)$, determined using all beads, at the location of the centroids only. 
As a result, the linear response approximation for the LMF-RPC charge density is
\begin{equation}
\rhoq(\rb) \approx \avg{\rhoq(\rb;\Rbar)}_0 -\beta\avg{\delta \rhoq(\rb;\Rbar) \delta {\Phi}^{\rm RPC}_{\rm R1}(\Rbar)}_0,
\label{eq:rpcrho}
\end{equation}
with an analogous expression for the centroid charge density.
The converged charge density and $\Vr(\rb)$ are obtained within the LMF-RPC approximation by iterating Eqs.~\ref{eq:rplmf} and~\ref{eq:rpcrho} to self consistency. 
%

\subsection{Centroid Approximation}
The LMF-RPC approach involves evaluating the charge density and centroid charge density, as well as the full $\Vr(\rb)$.
Such a complicated approach might not be necessary. 
Because $v_1(r)$ essentially smears the charge distribution over the length scale $\sigma$,
one might anticipate that a completely centroid approximation to the long range interactions should be reasonable when $\sigma$ is larger than the typical size of the ring polymers in the system. 
To determine the effective field via Eq.~\ref{eq:rplmf} within the centroid approximation,
the charge density operator appearing in Eq.~\ref{eq:qlrtq} is approximated everywhere
by the centroid charge density operator, $\rhoq(\rb;\Rbar)\approx \bar{\rho}^q(\rb;\Rbar)$. 
In this case, the centroid approximation to the LMF potential is $\Vr(\rb)\approx\bar{\V}_{\rm R}(\rb)$, where
\begin{align}
\bar{\V}_{\rm R}(\rb) &=\V(\rb) + \int d\rb' \bar{\rho}^q(\rb;\Rbar) v_1(\len{\rb-\rb'}) \\
&\equiv \V(\rb) + \bar{\V}_{\rm S}(\rb).
\end{align}
The centroid approximation is then inserted into Eq.~\ref{eq:qlrtq} to iterate the LMF equation to self-consistency
in conjunction with the linear response approximation for the centroid charge density
\begin{equation}
\bar{\rho}^q(\rb) = \avg{\bar{\rho}^q(\rb;\Rbar)}_0 -\beta\avg{\delta \bar{\rho}^q(\rb;\Rbar) \delta \bar{\Phi}_{\rm R1}(\Rbar)}_0,
\end{equation}
where $\bar{\Phi}_{\rm R1}(\Rbar)=\int d\rb \bar{\rho}^q(\rb;\Rbar) \bar{\V}_{\rm R1}(\rb)$ is the instantaneous energy
from the centroid approximation to the long range field evaluated at the centroid positions. 
%

\subsection{Feynman-Kleinert Approximation}

The LMF-RPC and centroid approaches reduce the number of sites needed to evaluate $\Vr(\rb)$, but
both still require a path integral simulation in the purely short range GT system to evaluate ensemble averages.
Instead, we could first model a classical ($P=1$) GT system and
use the Feynman-Kleinert (FK) procedure
to estimate the quantum LMF from its classical counterpart~\cite{feynman1986effective}. 
This second approximation, here called the FK approximation, in essence,
corresponds to approximating a quantum observable by a Gaussian smoothing of its classical counterpart~\cite{feynman1986effective}. 
This can be used to determine the LMF for the quantum system. 
First, I determine the LMF potential for the classical system, $\Vr^{\rm cl}(\rb)$,
by self-consistent iteration using the classical linear response approximation~\cite{Hu:PRL:2010}.
The linear response approximation is then used to predict the oxygen and hydrogen site densities
in the LMF system, $\rho_{\rm O}(\rb)$ and $\rho_{\rm H}(\rb)$, respectively. 
I then smooth these densities over the lengths $l_{\rm O}$ and $l_{\rm H}$ to convert the classical charge density into an approximation
of the quantum charge density,
\begin{align}
\rho^{ql}(\rb) &= \int d\rb' \big[q_{\rm O} \rho_{\rm O}(\rb') \rhoG(\len{\rb-\rb'};l_{\rm O}) \nonumber \\
&+ q_{\rm H} \rho_{\rm H}(\rb') \rhoG(\len{\rb-\rb'};l_{\rm H})\big],
\end{align}
where 
\begin{equation}
\rhoG(r;l)=\frac{1}{l^3 \pi^{3/2}} e^{-r^2/l^2}
\end{equation}
is a spherical Gaussian of width $l$.
Physically, $l$ corresponds to the average size of a ring polymer,
quantified by its radius of gyration, for example. 
Because of the different masses of oxygen and hydrogen, and consequently different spreads, we need different smoothing lengths for each,
$l_{\rm O}$ and $l_{\rm H}$, respectively, and
we need to separate the charge density into its components from oxygen and hydrogen sites.
Here, I approximate the size of the ring polymers by their
free particle values, $l_{\rm O}\approx \lambda_{\rm free, O}\approx0.05$~\AA \ and $l_{\rm H}\approx \lambda_{\rm free, H}\approx 0.2$~\AA.
This crude approximation is reasonable for water at ambient conditions, where the average radius of gyration of the ring polymers
is close to their free particle values~\cite{Markland:2008}.
After the densities are smoothed to account for nuclear quantum effects within the FK approximation,
I then perform a second smoothing over the length scale $\sigma$ by convoluting the quantum charge density with $v_1$.
The resulting FK approximation to the LMF potential is
$\Vr(\rb)\approx \Vr^{\rm FK}(\rb)$, where
\begin{align}
\Vr^{\rm FK}(\rb) &= \V(\rb) + \int d\rb' \rho^{ql}(\rb') v_1(\len{\rb-\rb'}) \\
&\equiv \V(\rb) + \V_{\rm S}^{\rm FK}(\rb).
\end{align}
In water at room temperature, the long range electrostatic interactions vary much more slowly
than the ring polymers for the nuclei, $\sigma\gg \l_{\rm H} > l_{\rm O}$. 
Therefore, one may anticipate that any deficiencies in the FK approximation --- deviations of the ring polymer
from a spherical Gaussian --- will be washed out by smoothing over $\sigma$,
and the FK approximation to the LMF potential will be reasonably accurate
at these conditions.
%

\begin{figure*}[tb]
\begin{center}
\includegraphics[width=0.95\textwidth]{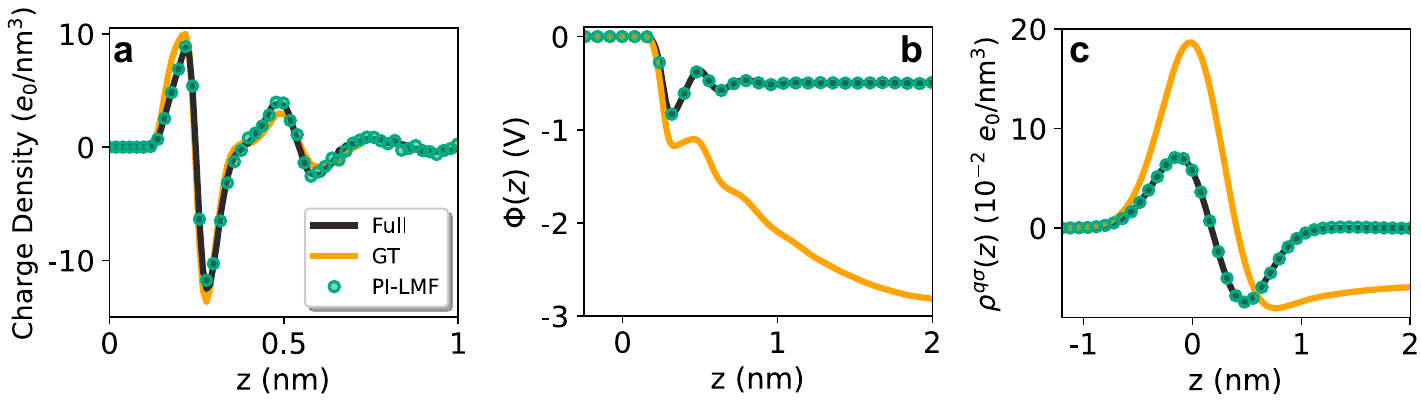}
\end{center}
\caption
{(a) Charge density, (b) Electrostatic potential, $\Phi(z)$, and (c) Gaussian smoothed charge density
for the full, Gaussian-truncated (GT), and PI-LMF mimic systems for $P=32$ beads. 
Results are shown for the left wall only. 
}
\label{fig:gt}
\end{figure*}

\section{Results and Discussion}
I demonstrate the utility of the LMF-RPC scheme using what
has become the canonical system for examining LMF theory-based methods --- 
water confined between idealized smooth hydrophobic walls. 
Near the walls, the effects of long range interactions do not cancel, as they do in bulk,
but instead create forces near the water-wall interface. 
The averaged effects of long range electrostatics provides a torque on interfacial water
molecules, resulting in dipolar screening
and the proper orientational ordering of water molecules near the wall~\cite{Rodgers:PNAS:2008,baker2020local,cox2020dielectric}.
Neglecting long range electrostatics in this system results in over-orientation of interfacial water molecules
and a non-zero electric field in the bulk due to the absence of dielectric screening in the purely short range system. 
Here, I show that the same general physics arises in path integral representations of water confined between hydrophobic walls
and that LMF theory adapted to path integral methods can account for the averaged effects of long range electrostatics. 

\subsection{Ring Polymer Local Molecular Field Theory can Account for Long Range Electrostatics}

The role of long range electrostatics in determining the structure of confined water can be observed
by comparing simulation results obtained with the Full and truncated models, Fig.~\ref{fig:gt}.
The charge density of water is similar in the Full, GT, and PI-LMF systems.
However, previous work has shown that differences relevant to long range electrostatics
are often hidden under large, atomic-scale fluctuations in the charge density.~\cite{Rodgers:PNAS:2008,LMFDeriv,remsing2015hydrophobicity}
These differences influence the orientational preferences of interfacial water.
In the GT systems, O-H bonds point toward the wall more than in the full systems, consistent with expectations from classical simulations. 
The orientation of interfacial water molecules alters the electrostatic potential, given by
\begin{equation}
\Phi(z) = -\int_{-\infty}^z dz' \int_{-\infty}^{z'} dz'' \rhoq(z'').
\end{equation}
The resulting electrostatic potential determined from GT configurations does not plateau in the bulk region. 
Including long range electrostatics, through Ewald summation (Full) or through the PI-LMF potential, corrects this behavior,
resulting in less orientational ordering of water at the interface and the expected plateau of the electrostatic potential in the bulk region.  

\begin{figure*}[tb]
\begin{center}
\includegraphics[width=0.95\textwidth]{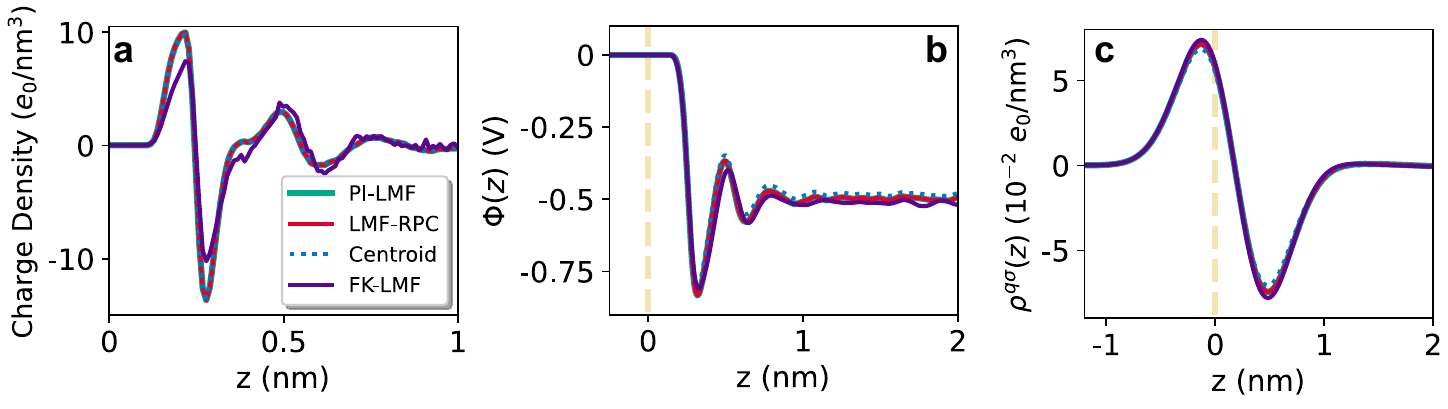}
\end{center}
\caption
{(a) Charge density, (b) Electrostatic potential, $\Phi(z)$, (c) Gaussian smoothed charge density,
and (d) gradient of the LMF potential (negative of force)
for the various methods of solving the LMF equation: the ``exact'' PI-LMF, LMF-RPC, the centroid approximation,
and the Feynman-Kleinert approximation (FK-LMF), all for $P=32$ beads. 
Results are shown for the left wall only. 
Vertical dashed line shows the location of the wall, $z=0$.
}
\label{fig:approx}
\end{figure*}

%
The long range part of the LMF potential satisfies a Poisson equation,
\begin{equation}
\nabla^2 \Vrl(\rb) = -\frac{4\pi}{\varepsilon} \rho^{q\sigma}(\rb),
\end{equation}
involving the Gaussian-smoothed charge density~\cite{LMFDeriv}
\begin{equation}
\rho^{q\sigma}(\rb)  = \int d\rb' \rhoq(\rb') \rho_G(\len{\rb-\rb'};\sigma),
\end{equation}
which is shown in Fig.~\ref{fig:gt}c for the Full, GT, and LMF systems. 
The Gaussian-smoothed charge density therefore represents the portion
of the charge density that is relevant for the long range response of water,
and one can think of $\rho^{q\sigma}(\rb)$ as a macroscopic charge density~\cite{robinson1973macroscopic,Zangwill,Remsing:JPCB:2015}.
The Full $\rhoqs(z)$ displays a (macroscopic) dipole layer at the interface.
In contrast, GT water overpolarizes and produces a large positive peak in $\rhoqs(z)$
at the interface.
Moreover, $\rhoqs(z)$ does not go to zero in the bulk, again due to overpolarization in the GT system.
The PI-LMF potential corrects this overpolarization and results
in a $\rhoqs(z)$ consistent with the dipole layer produced in the full system.  
This indicates that the PI-LMF system reproduces the long range behavior of confined quantum water.

In addition to the PI-LMF solution to the LMF equation, I described three approximate solutions to the LMF equation:
the LMF-RPC approximation, the centroid approximation, and the Feynman-Kleinert approximation (FK-LMF). 
The results obtained using these approximations are compared to the PI-LMF results in Fig.~\ref{fig:approx}. 
The charge densities agree for all systems but the FK approximation, which produces slightly smaller first peaks.
The discrepancy in the FK results stems from the inability of the linear response approximation to sufficiently shift the first peaks in the atomic densities; direct simulation of GT water in the presence of $\Vr^{\rm FK}(z)$ produces a charge
density in agreement with the others. 
Despite this difference in the results of the FK approximation, the electrostatic potentials and smoothed charge densities all agree,
indicating that these approximations for the long range interactions are reasonable for water at ambient conditions. 
This good agreement among the various approximations is a consequence of washing
out molecular-scale details in the coarse-graining inherent within the LMF approach.
However, one might expect that these approximations could break down when the thermal radius of the quantum particles
is comparable to or greater than the smoothing length, $\sigma$, as is the case for light nuclei at low temperatures~\cite{ceperley1995path,myung2022prediction},
or for light particles like electron or hole quasiparticles~\cite{Remsing_2020,park2022nonlocal}.
In cases like this, approximating a quantum particle by its centroid could prove inaccurate, especially if the ring polymers are aspherical. 

\subsection{Quantum Water Works Harder for Dipolar Screening}

\begin{figure}[tb]
\begin{center}
\includegraphics[width=0.48\textwidth]{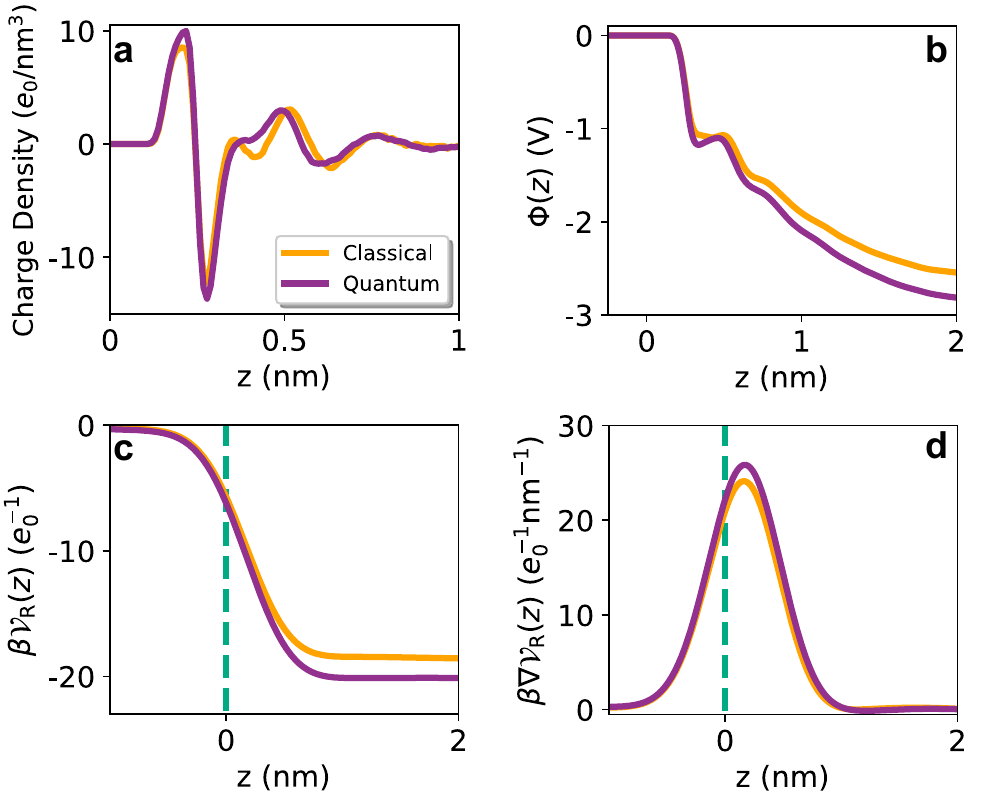}
\end{center}
\caption
{(a) The charge density and (b) polarization potential for the classical ($P=1$) and quantum ($P=32$) GT systems. 
The corresponding (c) LMF potential and (d) its gradient (negative of force) for the same two systems. 
Results are shown for the left wall only. 
}
\label{fig:gtcomp}
\end{figure}

The impact of NQEs on long range electrostatics can be assessed
by examining the extent to which the quantum and classical GT systems deviate
from our expectations for the full system.
The charge densities for the quantum and classical GT systems, shown in Fig.~\ref{fig:gtcomp}a, are qualitatively similar
but display small differences near the wall. 
In particular, the magnitudes of the first two peaks are slightly larger for the quantum GT model. 
These small differences between the charge densities lead to large differences when integrated
to obtain the polarization potential, as shown in Fig.~\ref{fig:gtcomp}b. 
The potential in the bulk region of the quantum system is significantly larger than that of the classical one. 
However, the differences in the electric fields (derivative of the potential) are localized near the interface,
and the fields are similar in the bulk for both quantum and classical systems. 
The overpolarization of interfacial water results from the lack of long range forces that reorient molecules near the wall. 
The LMF potential provides these long range forces and corrects interfacial water structure. 
%

%
Because quantum GT water overpolarizes more than classical GT water,
the forces needed to reorient quantum GT water should be larger. 
Indeed, the corresponding LMF potentials and forces are larger in quantum GT water due to NQEs, Fig.~\ref{fig:gtcomp}c,d.
The LMF, $\Vr(z)$, exhibits a larger change across the interface for the quantum systems at all levels of approximation. 
Moreover, the LMF force, $-\nabla \Vr(\rb)$, is larger in magnitude for the quantum systems.
This suggests that the forces required to achieve proper dielectric screening are larger in quantum systems than their classical counterparts,
by several $\kT$ for water at room temperature,
which may be anticipated from the larger zero point energy of the quantized nuclei. 

\subsection{Computational Efficiency}

\begin{figure}[tb]
\begin{center}
\includegraphics[width=0.48\textwidth]{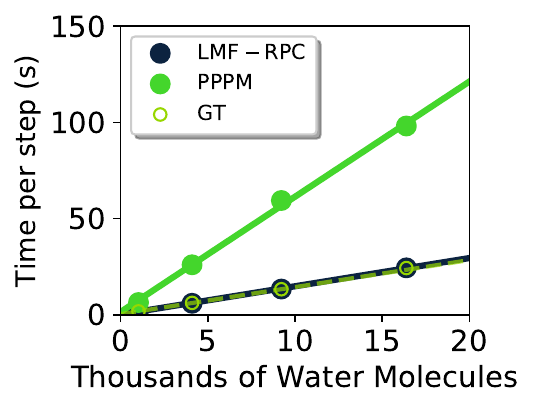}
\end{center}
\caption
{Computation time per MD simulation time step as a function of the number of water molecules in the system. 
Lines are linear fits to the data. Error bars are smaller than the symbol size.
}
\label{fig:time}
\end{figure}

Most of the results above were obtained using linear response theory and simulations of
a short range system. 
However, simulations can be performed with the presence of the LMF field
to also obtain accurate predictions of the structure of nonuniform fluids. 
Compared to typical lattice summation techniques
for evaluating long range electrostatics, the LMF-RPC scheme reduces the cost significantly in two ways.
First, RPC reduces the number of sites at which long range interactions need to be evaluated --- here,
from 32 to 1. 
Second, LMF theory replaces $N^2$ two-body interactions with $N$ one-body interactions.
This reduction in scaling is beneficial when simulating large numbers of molecules characteristic of
biological and materials systems. 
To illustrate the increased efficiency of the LMF-RPC approach in comparison to Ewald summation-based approaches
for evaluating long range electrostatic interactions in path integral models,
I evaluated the time required to perform a PIMD time step as a function of the number of water molecules in the system,
shown in Fig.~\ref{fig:time}.
The number of water molecules was varied by replicating the simulation box in the lateral directions ($x$ and $y$). 
The computational time was evaluated for the LMF-RPC, particle-particle-particle mesh (PPPM) Ewald,
and GT systems with $P=32$ using a single 2.3 GHz Intel Xeon W core of a 2017 iMac Pro.
For the small system size used in the majority of the text (1024~molecules), the simulation time is similar for all approaches. 
However, as the size of the system grows, the increased efficiency of the LMF-based approaches becomes apparent. 
Moreover, the timings for the GT system are nearly identical to the LMF systems, indicating
that the evaluation of the (one-body) LMF potential requires minimal overhead and the calculation
is dominated by the evaluation of short range pairwise interactions. 
Of course, this means that there is negligible speedup gained by using RPC with LMF theory for the system
sizes studied here, but differences may appear for systems with large numbers of beads. 
In contrast to the LMF results,
the PPPM timings are slowed by the evaluation of the long range electrostatic interactions. 
This suggests that LMF-based approaches can drastically reduce the computational cost of PIMD calculations
in large-scale molecular systems. 
%

\section{Conclusions}
In this work, I have examined nuclear quantum effects on long range electrostatic properties of confined water. 
To do so, I demonstrated that LMF theory can be used to efficiently and accurately
account for long range electrostatics in
path integral simulations of nonuniform liquids. 
Moreover, a RPC-based approximations were introduced that leverage the complementary ideas that underlie the separation
of short and long range interactions in both LMF theory and RPC --- long range interactions are slowly varying 
over relevant molecular length scales.
I expect that the LMF-RPC scheme will be useful for modeling NQEs in large systems with many light particles
at low temperatures (many beads).
The LMF-RPC scheme can be readily combined
with developments in LMF theory to evaluate NQEs on free energy differences~\cite{Remsing:2016ib,Remsing:JTCC:2018}. 
The general ideas presented here may also be valuable for modeling NQEs with the symmetry preserving
mean field (SPMF) theory, which replaces pairwise long range electrostatics with a symmetry-dependent 
effective field in each configuration~\cite{Hu2014,Yi2017}.
The LMF-RPC approach may be particularly powerful when combined with the short solvent model of
Gao~\etal~\cite{SSM,gao2023local} for molecular assembly. 
In the short solvent model, long range solvent-solvent and solute-solvent interactions are truncated everywhere,
and the averaged effects of all long range interactions are accounted for with effective solute-solute interactions.
Therefore, the only possible long range interactions are between solutes,
greatly reducing the number of charged sites
and the associated cost of evaluating long range interactions. 
Combining the short solvent model with RPC will result in a model where long range interactions
only need to be evaluated
for the solute centroids. 
All other interactions are short range. 
The resulting SSM-RPC scheme could be of great use for modeling NQEs in self-assembly processes. 
The results reported here use empirical force fields to represent intra- and intermolecular interactions,
which neglect coupling of electronic polarization to nuclear quantum fluctuations. 
These effects can be taken into account with ab initio simulations.
Previous work has shown that RPC can significantly speed up ab initio simulations by using
a cheap approach to evaluate interactions on all beads, e.g. density functional tight binding (DFTB) methods,
while higher level ab initio
methods are evaluated only on the centroid and the resulting bead-bead interactions are obtained 
from the differences between these two levels of theory~\cite{marsalek2016ab,marsalek2017quantum}. 
Such an approach does not readily lend itself to treatment with LMF theory. 
However, if LMF theory or similar approaches can be extended to ab initio models,
this would facilitate the use of the LMF-RPC scheme in ab initio path integral simulations. 
An alternative to costly ab initio simulations is to use machine learning approaches to develop neural network potentials
that can produce ab initio accuracy with much smaller cost~\cite{Zhang2018,Behler2021,batzner20223}.
Traditional neural network potentials lack a good description of long range interactions~\cite{yue2021short},
but recent developments include some description of long electrostatics~\cite{ko2021fourth,grisafi2019incorporating,yue2021short,niblett2021learning,Gao:NatureComm:2022,Dhattarwal:JPCB:2023}. 
Of particular interest are neural network potentials that are informed LMF ideas~\cite{niblett2021learning,Gao:NatureComm:2022,Dhattarwal:JPCB:2023},
like the self-consistent field neural network (SCFNN)~\cite{Gao:NatureComm:2022,Dhattarwal:JPCB:2023}.
These networks focus on training short range GT interactions separately from long range interactions. 
I anticipate that many of these neural network potentials could be combined with the LMF-RPC scheme to treat the averaged effects of long range interactions in path integral and ring polymer MD simulations with ab initio accuracy.
%

\section{Simulation Details}
Path integral molecular dynamics (PIMD) simulations were performed using the i-PI package~\cite{kapil2019pi} interfaced
with the LAMMPS software package~\cite{LAMMPS}, modified to include the truncated potential $v_0(r)$ and the LMF potential $\Vr(z)$. 
Equations of motion were integrated in the normal mode representation using the Cayley integration scheme with a timestep of 0.5~fs~\cite{korol2019cayley}.
All simulations were performed in the canonical (NVT) ensemble
at a constant temperature of 298~K maintained using a stochastic velocity rescaling thermostat~\cite{bussi2007canonical},
with 1024 water molecules in a $27.72\times27.72\times150.00$~\AA$^3$ simulation cell. 
All simulations employed the flexible quantum SPC water model of Voth and coworkers~\cite{paesani2006accurate}.
Lennard-Jones and short range Coulomb interactions we truncated at a distance of 9.8~\AA.
Idealized hydrophobic walls were each represented with a 9-3 Lennard-Jones potential,
\begin{equation}
U_{\rm w}(z) = \varepsilon_{\rm w}\brac{\frac{2}{15}\para{\frac{\sigma_{\rm w}}{z}}^9 - \para{\frac{\sigma_{\rm w}}{z}}^3},
\end{equation}
with $\sigma_{\rm w}=3.461$~\AA \ and $\varepsilon_{\rm w}=0.43875491$~kcal/mol.
Wall-water interactions are cut off at a distance of 30~\AA.
Walls are positioned at $z=0$ and $z=43.06$~\AA, in accord with previous work~\cite{Rodgers:PNAS:2008}.
All ring polymers used $P=32$ beads for each particle in the system, which has been shown to be sufficiently
converged~\cite{Markland:2008}.
The modified LAMMPS source code is available at github.com/remsing-group/lmf-rpc/. 

\vspace{1cm}
\acknowledgements
This work is supported by the National Aeronautics and Space Administration under grant number 80NSSC20K0609, issued through the NASA Exobiology Program.
I acknowledge the Office of Advanced Research Computing (OARC) at Rutgers,
The State University of New Jersey
for providing access to the Amarel cluster 
and associated research computing resources that have contributed to some of the results reported here.
I thank Atul Thakur for helpful comments on the manuscript
and D. Rodman for inspiration on the figures. 

\bibliography{Refs}

\end{document}